# High-efficiency Ventilated Metamaterial Absorber at Low Frequency

Xiaoxiao Wu[1,a], Ka Yan Au-Yeung[1,a], Xin Li[2], Robert Christopher Roberts[3], Jingxuan Tian[3], Chuandeng Hu[1], Yingzhou Huang[2], Shuxia Wang[2], Zhiyu Yang[1], Weijia Wen[1,b]

[1]*Department of Physics, The Hong Kong University of Science and Technology, Clear Water Bay, Kowloon, Hong Kong, China*

[2]*Department of Applied Physics, Chongqing University, Chongqing 401331, China*

[3]*Department of Mechanical Engineering, Faculty of Engineering, The University of Hong Kong, Hong Kong, China*

[a] Xiaoxiao Wu and Ka Yan Au-Yeung contributed equally to this work.

[b] To whom correspondence should be addressed. Electronic mail: phwen@ust.hk

**ABSTRACT**

We demonstrate a ventilated metamaterial absorber operating at low frequency (< 500 Hz).With only two layers of the absorption units, high-efficiency absorption (> 90%) has been achieved in both simulations and experiments. This high-efficiency absorption under ventilation condition is originated from the weak coupling of the two identical split tube resonators constituting the absorber, which leads to the hybridization of the degenerate eigenmodes and breaks the absorption upper limit of 50% for conventional transmissive symmetric acoustic absorbers. The absorber can also be extended to an array and work in free space. The absorber should have potential applications in acoustic engineering where both noise reduction and ventilation are required.

Absorbing low-frequency (< 500 Hz) sounds is a long-standing challenge in acoustic researches. Conventional porous materials could not achieve high-efficiency absorption without a thickness comparable to the wavelength.[1,2] To increase the absorption performance, we need to increase the density of states of acoustic waves at low frequency. Therefore, artificial structures, collectively called "metamaterials", becomes popular since they can support deep-subwavelength resonance modes and offer large density of states at low frequencies. Under this concept, many high-efficiency and subwavelength acoustic metamaterial absorbers have been proposed and realized over the last few years.[3-9] Remarkably, an optimal acoustic metamaterial absorber has also been demonstrated.[10] A common feature of them is that they all require reflectors, such as aluminum plates or rigid walls, placed behind them to prevent transmission of sounds. If the reflectors behind these absorbers are removed, their maximum absorptions will be limited to only 50%.[11,12] However, fluid flows, such as air or water, will also be totally prevented from circulating by these reflectors. To break the absorption limit and meanwhile maintain no reflectors to allow ventilation, several coherent perfect absorbers (CPAs), originally adopted in optics as the time-reversal counterpart of lasers,[13,14] have been proposed and realized in acoustics.[15-17] However, the dynamic generation and real-time control of an extra incident sound beam is not easy and can be expensive in real applications, thus a passive high-efficiency ventilated absorber is still a more practical option. Consequently, several ventilated absorbers have been proposed to meet this demand,[18-22] but their high-efficiency absorptions are only for one side, while for the other side the absorptions are usually minimized. Yang *et al*. has realized a ventilated perfect absorber whose absorption is identical for both sides,[23] but the design requires membranes, of which the prestress that determines the resonance frequency of the absorber is difficult to precisely control; the absorption bandwidth is also very limited (< 3Hz). Therefore, it is desirable to develop a subwavelength and tunable ventilated absorber with high-efficiency absorption for both sides. Such ventilated absorber may further cooperate with recently proposed ventilated reflectors[24,25] to create a pleasant sound environment while keeping a good ventilation.

In this Letter, we demonstrate a ventilated metamaterial absorber (VMA) which can achieve high-efficiency (> 90%) absorption for one-sided incidence. As the name indicates, the VMA permits the flow of fluids from both sides since it does not require any reflectors. Moreover, the VMA can work in both waveguides and free space, which suggests a variety of working scenarios.

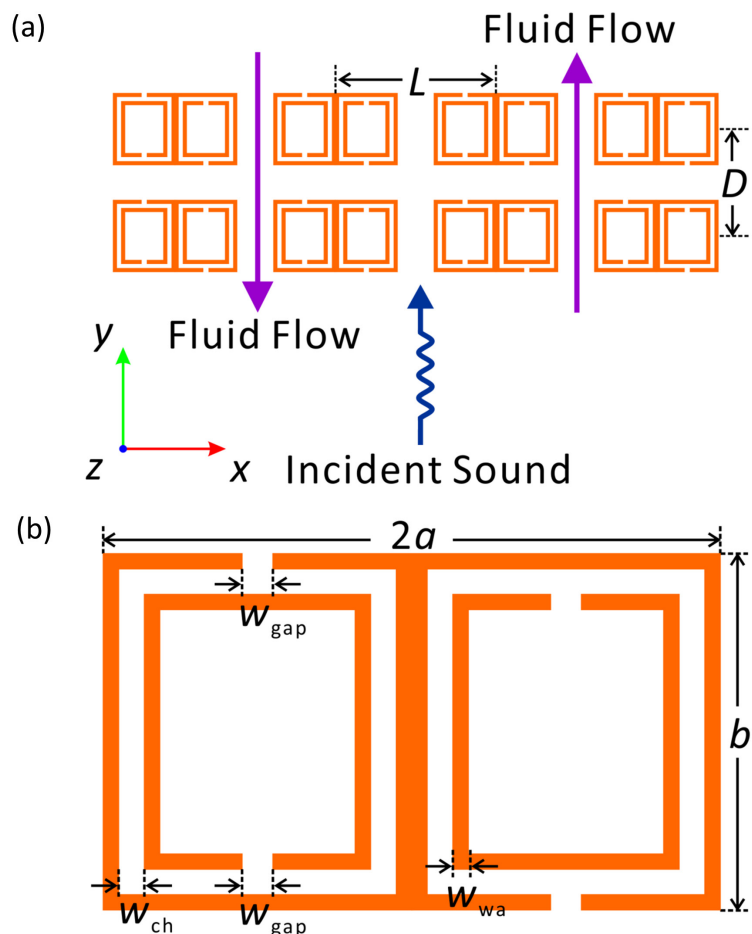

FIG. 1. Geometry of the designed ventilated metamaterial absorber (VMA). (a) Schematic cross section illustrating the VMA composed of several layers of the absorption units which are invariant along the $z$ direction. The dashed rectangle denote an absorption unit of the VMA. (b) Details of an absorption unit of the VMA, which is composed of two identical but oppositely oriented split tube resonators. A chosen set of the geometric parameters for demonstration purpose is listed in Table I.

As depicted in Fig. 1(a), the designed VMA is composed of several layers of absorption units (two such layers depicted in Fig. 1(a)) arranged in a one-dimensional array along the $x$ direction. The gaps between the absorption units permit the flows of various fluids, such as air or water, passing through the absorber. The details of an absorption unit are shown in Fig. 1(b), which comprises two identical but oppositely oriented split tube resonators,[8] invariant along the $z$ direction. For demonstration purpose, we assume that the VMA is immersed in air and choose a set of the geometric parameters for the VMA and they are listed in Table I.

TABLE I. Geometric parameters of the designed VMA. All parameters are in unit of mm.

| $L$ | $a$ | $b$ | $w_{ch}$ | $w_{gap}$ | $w_{wa}$ |
|---|---|---|---|---|---|
| 90 | 35 | 45 | 1.6 | 1.6 | 2 |

We then examine the acoustic performance of VMAs using COMSOL Multiphysics. The setup of simulations is detailed in supplementary material. Initially, we consider a VMA comprising only one layer of the absorption units. The complex transmission coefficient $t$ and reflection coefficient $r$ are retrieved from simulations and the absorption coefficient $A$ then is calculated as $A = 1-T-R$ in which $T = |t|^2$ and $R = |r|^2$ because of the conservation of energy. Since the structure possesses spatial inversion symmetry, the responses to the wave propagating along $+y$ or $-y$ directions are identical and hence we do not differentiate them (see supplementary material). The simulated results are shown in Fig. 2(a), and a notable absorption (88.9%) is achieved with only one layer of the absorption units at the frequency 343 Hz indicated by the black arrow. Similar to previous works[5,8], the energy dissipation is due to the strong friction at the narrow necks of the split tube resonators caused by the pressure difference between the interior cavities and outside environment.

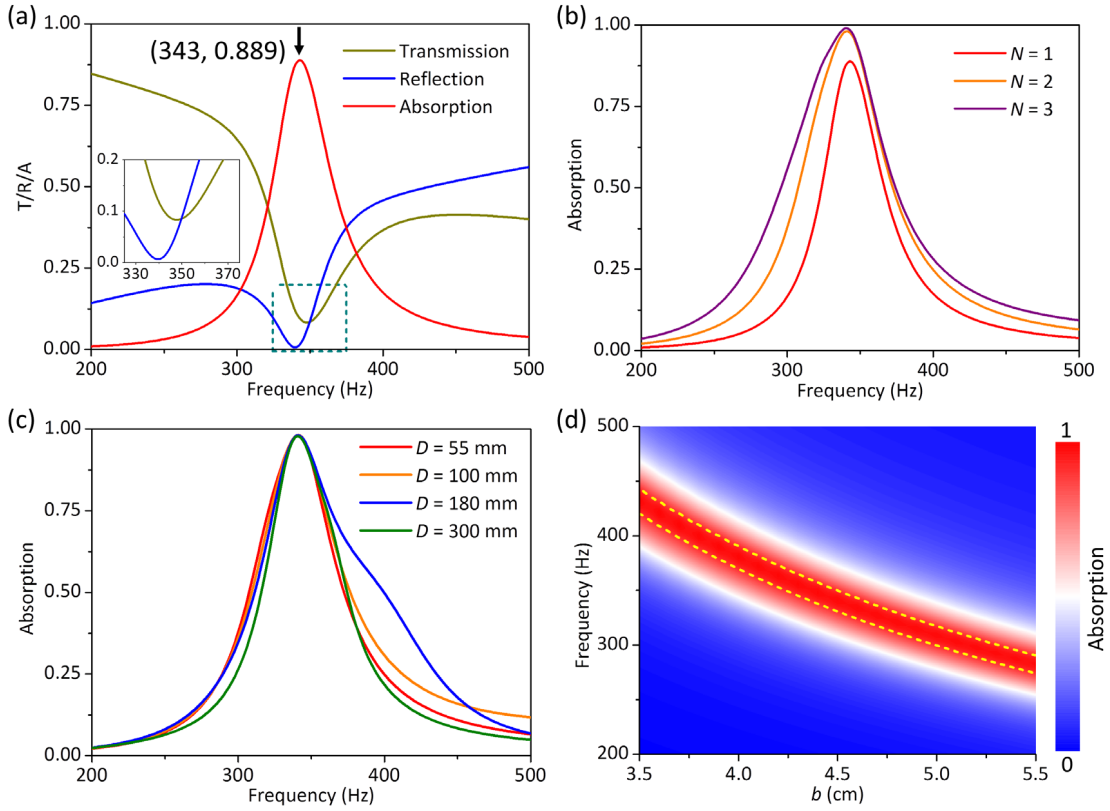

FIG. 2. Simulated acoustic performance of the VMA. (a) Simulated transmission($T$)/reflection($R$)/absorption($A$) of the VMA composed of a single layer. The black arrow denotes the maximum absorption 88.9 % at 343 Hz (b) Simulated absorption of the VMA composed of $N$ ($N$ = 1, 2, 3) layers. $D$ = 55 mm. (c) Simulated absorption of the VMA composed of two layers with varying distance *D*. (d) Simulated absorption of the VMA composed of two layers of the units with varying thickness *b*. The dashed yellow lines denote the contour of 90% absorption.

Near maximum absorption, we can observe that the reflection is very small (< 0.7%), but the transmission is not negligible (~ 8.3%), as displayed in the inset of Fig. 2(a). This small reflection reminds us that we can further increase the absorption by directly adding more layers of the absorption units. To verify this deduction, we simulated the absorption of the VMA composed of more layers. The simulated results are summarized in Fig. 2(b) and it can be seen that a near-unity (> 98%) absorption has been achieved when there are only two layers of such absorption units. When there are three layers, an absorption > 99% is achieved. Because the absorption is due to the localized mode of the split tube resonators, the distance between each layer should not substantially affect the absorption performance. For proof, the simulated absorptions of the two layers with considerably different center-to-center distances *D* are shown in Fig. 2(c) and there are no remarkable differences between these results, especially around the maximum absorption frequency. By adjusting the geometric parameters, the high-efficiency absorption can be shifted to other frequencies. For example, we consider a VMA composed of two layers of the absorption units and adjust the thickness of absorption unit *b*, while other geometric parameters are identical with those in Table I. The simulated absorption is plotted as a contour map against the thickness *b* and the frequency in Fig. 2(d), and 90% absorption is denoted by the yellow dashed lines. As can be seen from the contour map, the high-efficiency absorption is robust when we adjust the thickness of the absorption units. The maximum absorption frequency is shifted but the maximum absorption is larger than 90%, which suggests that we can customize the absorption units for a certain frequency. We also calculate the flow resistivity of the VMA in the waveguide and free space, and the results show that the effective flow resistivity (<10 Pa·s·m$^{-2}$) is much smaller compared with conventional porous materials (2000-40000 Pa·s·m$^{-2}$)[26] and hence the VMA will not significantly impede the air flow (see supplementary material).

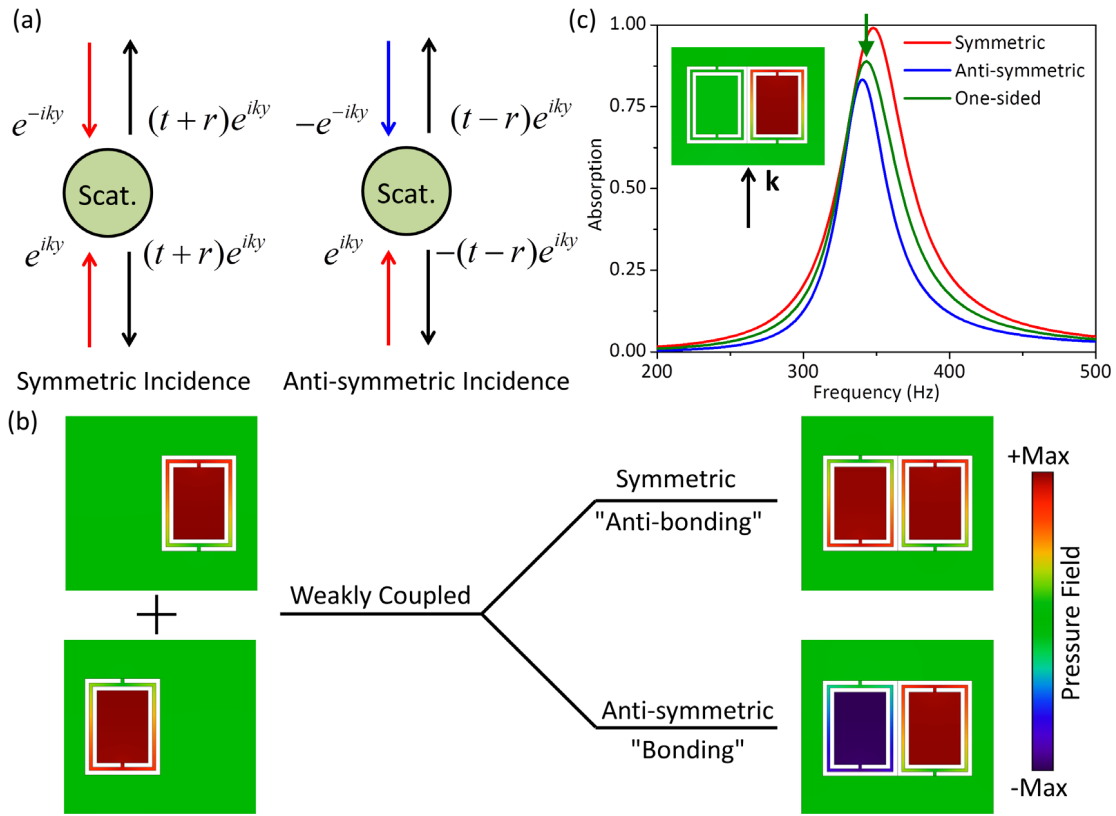

FIG. 3. Origin of high-efficiency absorption of the VMA under one-sided incidence. (a) Schematic sketches illustrating the situation of (left) symmetric incidence and (right) anti-symmetric incidence. "Scat." stands for "Scatter". (b) Diagram of the frequency split of the two weakly coupled split tube resonators. The degenerate eigenmodes of the two split tube resonators are split into a symmetric mode with a higher frequency ("Anti-bonding") 346 Hz and an anti-symmetric mode with a lower frequency ("Bonding") 339 Hz. (c) Simulated

symmetric absorption $A_s$, anti-symmetric absorption $A_a$, and one-sided absorption $A$ of the VMA. The inset shows the snapshot of the pressure field of the VMA at 343 Hz as indicated by the green arrow under one-sided incidence as denoted by the black arrow.

In order to understand why the subwavelength VMA with inversion symmetry can achieve an absorption notably larger than 50% without reflectors, we first reveal why conventional subwavelength absorbers with inversion or mirror symmetry cannot achieve this performance. For convenience, we consider a reciprocal scattering process in an one-dimensional system with two ports and assume that the scatter is symmetric with respect to the two ports. In such configuration, the scattering process could be described using a scattering matrix[19] $\mathbf{S}(f)$ which relates the phasors of the incoming waves $[a, b]^T$ to the phasors of the outgoing waves $[c, d]^T$

$$\begin{bmatrix} c \\ d \end{bmatrix} = \mathbf{S}(f) \begin{bmatrix} a \\ b \end{bmatrix} = \begin{bmatrix} t & r \\ r & t \end{bmatrix} \begin{bmatrix} a \\ b \end{bmatrix}. \tag{1}$$

We can simply verify that the eigenvalues of $\mathbf{S}(f)$ are $t \pm r$, which corresponds to a symmetric incidence $[1, 1]^T$ or an anti-symmetric incidence $[1, -1]^T$, respectively, and they are sketched in Fig. 3(a). We can then decompose the one-sided incidence $[1, 0]^T$ into a superposition of the symmetric incidence and anti-symmetric incidence.[15] This decomposition inspires us to rewrite the expression of the absorption $A$ as[19]

$$A = 1 - |r|^2 - |t|^2 = \frac{1-|r+t|^2 + 1-|r-t|^2}{2} = \frac{1}{2}(A_s + A_a), \tag{2}$$

in which the second equal sign is due to the identity $|r+t|^2+|r-t|^2 = 2(|r|^2+|t|^2)$. Further, as sketched in Fig. 3(a), $A_s = 1-|r+t|^2$ is the absorption under symmetric incidence, while $A_a = 1-|r-t|^2$ is the absorption under anti-symmetric incidence. We simply call $A_s$ and $A_a$ as symmetric and anti-symmetric absorption, respectively. Because of the conservation of energy, for a passive scatter, we have the following constraints

$$0 \leq A_a \leq 1, \tag{3}$$

and

$$0 \leq A_s \leq 1. \tag{4}$$

When the scatter has an inversion center or a mirror axis perpendicular to the wave propagating direction, its eigenmodes will be symmetric or anti-symmetric under the symmetry operation. It can be proven (see supplementary material) that for subwavelength scatters, a symmetric (anti-symmetric) eigenmode is transparent for anti-symmetric (symmetric) incidence. If the subwavelength scatter only have a symmetric or anti-symmetric eigenmode in the considered frequency range, only its symmetric absorption $A_s$ or anti-symmetric absorption $A_a$ will be nonzero, respectively. As a direct result of Eqs. (3) and (4), its one-sided absorption $A = A_s/2$ or $A = A_a/2$ then cannot exceed 50% in this frequency range. Similar conclusions have been proven for several specific acoustic absorbing structure in previous works.[12,19]

However, because here we have two weakly coupled identical split tube resonators, their degenerate eigenmodes will interact with each other and be split into an anti-symmetric mode with a lower eigenfrequency ("Bonding") at 339 Hz and a symmetric mode with a higher eigenfrequency ("Anti-bonding") at 346 Hz, as shown in Fig. 3(b). We further extract the symmetric absorption $A_s$ and anti-symmetric absorption $A_a$ of the VMA from simulations, and plot them as well as the one-sided absorption $A$ in Fig. 3(c). It can be observed that both symmetric and anti-symmetric eigenmodes are well excited under two-sided incidence and these two absorption peaks are very close and overlap with each other. The coexistence of large symmetric and anti-symmetric absorptions then ensures that our absorber can achieve a high-efficiency absorption at the frequency (343 Hz) between two eigenmodes. The inset shows the snapshot of the pressure field around the VMA at 343 Hz when the wave is incident from $-y$ direction and it can be observed that the pressure in the left split tube resonator is very small, indicating that the excited field inside the VMA under one-side incidence is indeed a linear superposition of the symmetric and anti-symmetric modes.

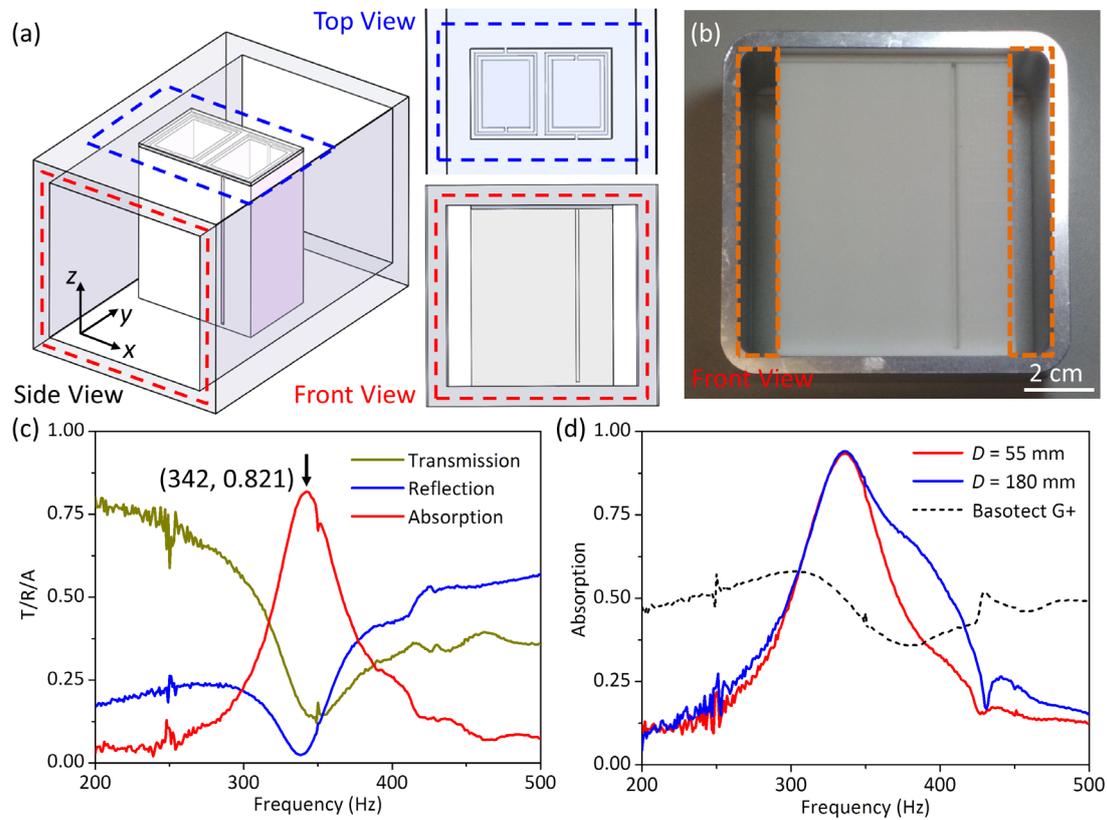

FIG. 4. Experimental characterization of the VMA. (a) Side-view, top-view, and front-view schematics of an absorption unit of the VMA placed in the impedance tube. Under normal plane-wave incidence, an absorption unit is equivalent to a layer of the absorption units extended along $x$ direction depicted In Fig. 1(a). (b) Front-view photograph of a fabricated sample made from polylactide (PLA) plastic placed in a section of the impedance tube.

Dashed orange rectangles denote the gap between the sample and the impedance tube. (c) Measured transmission(*T*)/reflection(*R*)/absorption(*A*) of a single absorption unit. The black arrow denotes the maximum absorption 82.1% at 342 Hz. (d) Measured absorption *A* of the VMA comprising two absorption units with varying distance *D*. The maximum absorption reaches 93.4% (*D* = 55 mm or 180 mm). The dashed line denotes the absorption of the reference sample, which is a melamine foam (Basotect G+) with thickness 100 mm, width 70 mm, and height 90 mm.

After understanding its working mechanism, to experimentally confirm the VMA, we measure the acoustic performance of its absorption units under normal incidence in a square impedance tube.[8] The side length of the cross section of the square impedance tube is 90 mm, corresponding to a plane wave cutoff frequency ~1900 Hz. We use the standard four-microphone method to measure the complex transmission coefficient *t* and reflection coefficient *r*. It should be noted that under normal plane-wave incidence, an absorption unit in the impedance tube is equivalent to a layer of absorption units extended along *x* direction with period *L* = 90 mm depicted in Fig. 1(a). This equivalence is because the wavelength (> 600 mm) is much larger than the period (90 mm) along *x* direction and hence there is no higher-order diffractions at low frequency (< 500 Hz).[8,27] For brevity, the detailed schematic diagram of the experiment setup is shown in supplementary material. In experiments, samples are placed in the middle of the impedance tube, as indicated in Fig. 4(a). These samples are made from polylactide (PLA) plastic by 3D printing techniques. A front-view photograph of the sample placed in a section of the impedance tube is shown in Fig. 4(b), in which the orange dashed rectangles denote the air gap between the sample and the impedance tube. In a series of experiments, we first measure a single absorption unit and the results are shown in Fig. 4(c), the maximum absorption 82.1% at 342 Hz denoted. It is observed that the measured results agree well with the simulated results in Fig. 2(a), except the maximum absorption is slightly lower than that in simulations. This discrepancy could be attributed to the fact that the PLA plastic we use to print the samples are not completely rigid which we assume in simulations. The assumption overestimates the pressure difference between inside cavities and outside environment, which in turn leads to the slight overestimation of the absorption performance in simulations. Nevertheless, the measured results confirm that the VMA can achieve absorption notably higher than 50%. We then measure the performance of two absorption units with the center-to-center distance $D$ = 55, 180 mm, respectively, and the results are plotted in Fig. 4(d). Both measured results again agree well with the corresponding simulated results shown in Fig. 2(c), confirming that adding more layers is feasible for improving the absorption of the VMA. For reference, we have also measured the absorption of a commercial melamine foam (Basotect G+) with its thickness, width, and height being 100 mm, 70 mm, 90 mm, respectively, using the same experiment setup, and the result is plotted as the black dashed line in Fig. 4(d). It is then apparent that the VMA is more efficient in a broadband (> 80 Hz) near resonance compared with the melamine foam as expected.

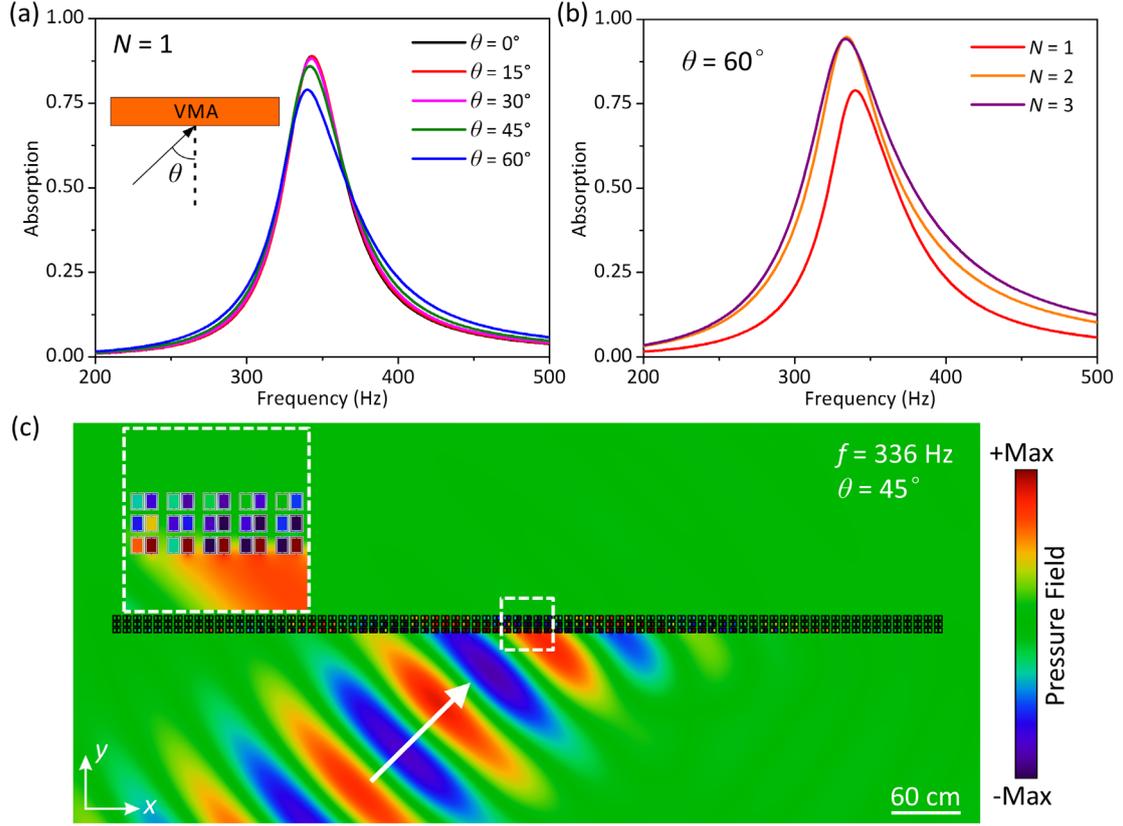

FIG. 5. Performance of VMAs under oblique incidence in free space. (a) Simulated absorption of a single layer under oblique incidence. The inset defines the angle of incidence $\theta$. (b) Simulated absorption of the VMA comprising $N$ ($N$ = 1, 2, 3) layers of the absorption units. The distance between each layer is $D$ = 55 mm. (c) Simulated pressure field at 336 Hz when a two-dimensional Gaussian beam impinges on a VMA comprising three layers of absorption units in free space. The distance between each layer is $D$ = 55 mm. The propagating direction of the beam is indicated by the white arrow and the angle of incidence is 45°. The inset shows the enlarged field distribution of the region enclosed by the white dashed rectangle.

As we have experimentally validated the VMA, we further estimate its performance under oblique incidence which is important for applications in free space. We plot the simulated absorptions of a single layer under various angles of incidence $\theta$ in Fig. 5(a), with $\theta$ defined in the inset. It can be seen that there is no significant degradation of absorption performance under oblique incidence with $\theta$ up to 60°. We also calculate the absorption of two layers of the absorption units under oblique incidence and the results shown in Fig. 5(b) also suggest that the VMAs can maintain a satisfying performance under oblique incidence in free space. Further, to directly visualize its performance in free space, we consider a two-dimensional Gaussian beam of acoustic waves impinging on a VMA composed of 3 layers with $D$ = 55 mm. The waist radius of the beam is 1 m and the angle of incidence is 45°. A snapshot of the simulated pressure field at the frequency 336 Hz is shown in Fig. 5(c). It can be seen that no notable reflection or transmission is generated after the beam encounters the VMA, which indicate a high-efficiency absorption of the incident beam at this frequency. At lower frequencies, the VMA is mainly transmissive, while at higher frequencies, the VMA is mainly reflective, as indicated in Fig. 2(a) (see also supplementary material). We also consider the absorption of a VMA placed in the impedance tube under oblique incidence and the calculated results show that the VMA can also achieve high-efficiency absorption when the angle of incidence is as large as 45° (see supplementary material). The robust performance suggests that when applying the VMA in a tube or duct, a small-angle (< 30°) misalignment error will not notably affect its absorption.

Finally, we point out that though we only demonstrate VMAs for audible sound in air, the design can be modified for applications in other frequency regimes and fluids, such as sonar sound (kHz) in water (see supplementary material), only requiring that it is made of materials rigid enough compared with the ambient environment. For example, when applied in water, the VMA should be made of brass or stainless steel, rather than plastics.

To summarize, we have demonstrated a ventilated acoustic absorber which can achieve high-efficiency absorption while maintaining a subwavelength thickness. The working frequency and absorption performance of the absorber is largely independent of its material properties and can be tuned by adjusting its geometry parameters. The absorber should be useful for applications which require both noise reduction and ventilation, such as in air conditioners and kitchen ventilators.

**SUPPLEMENTARY MATERIAL**

See supplementary material for setup of simulations in COMSOL Multiphysics, confirmation of symmetric acoustic performance of the VMA, flow resistivity of the VMA of different sizes, transparency of eigenmodes under symmetric or anti-symmetric incidence, schematic diagram of experimental setup, pressure fields of the VMA under oblique incidence in free space at other frequencies, absorption under oblique incidence in the impedance tube, and realization of the VMA in other working frequency regimes and fluids.

**ACKNOWLEDGEMENT**

The project is supported by an Areas of Excellence Scheme grant (AOE/P-02/12) from research grant council (RGC) of Hong Kong and the Special Fund for Agro-scientific Research in the Public Interest from Ministry of Agriculture of the Peoples' Republic of China (no. 201303045).